\documentclass[aps,prl,twocolumn,groupaddress]{revtex4}
\usepackage{amsmath}
\usepackage{amsfonts}
\usepackage{amssymb}
\usepackage{epsfig}
\usepackage{graphicx}

\usepackage[usenames,dvipsnames]{color}

\begin{document}

\title{Non-linear dynamic response of glass-forming liquids to random pinning}

\author{Walter Kob}
\affiliation{Laboratoire Charles Coulomb, UMR 5221, CNRS and Universit\'e
Montpellier 2, Montpellier, France}

\author{Daniele Coslovich}
\affiliation{Laboratoire Charles Coulomb, UMR 5221, CNRS and Universit\'e
Montpellier 2, Montpellier, France}

\begin{abstract}
We use large scale computer simulations of a glass-forming liquid in
which a fraction $c$ of the particles has been permanently pinned. We
find that the relaxation dynamics shows an exponential dependence on $c$.
This result can be rationalized by means of a simple theoretical Ansatz
and we discuss its implication for thermodynamic theories for the
glass-transition.  For intermediate and low temperatures we find that
the slowing down of the dynamics due to the pinning saturates and that
the cooperativity decreases with increasing $c$, results which indicate
that in glass-forming liquids there is a dynamic crossover at which the
shape of the relaxing entities changes.

\end{abstract}

\maketitle

The extensive studies done during the last two decades on the
relaxation dynamics of glass-forming liquids have shown that this
dynamics is intimately related to a cooperative motion of the
particles~\cite{berthier_11}.  In particular it has been found
that the number of particles involved in this dynamics increases
with decreasing temperature $T$~\cite{berthier_05,dalle_07}, thus
rationalizing the super-Arrhenius temperature-dependence of the relaxation
times~\cite{binder_11}. Although these results seem to confirm the old
ideas of Adam and Gibbs on the existence of cooperatively rearranging
regions (CRRs)~\cite{adam_65}, other theoretical approaches are compatible
with these findings as well~\cite{kirkpatrick_89,lubchenko_07,chandler_10}
and hence the question which theoretical description is the right one
is still open~\cite{binder_11,berthier_biroli_11}.

Usually this growing cooperativity is expressed via a dynamic
four-point correlation function and its associated length
scale~\cite{berthier_11,dalle_07,bennemann_99}.  Recently, however,
evidence has been given that also the {\it structure} of the CRRs may
depend on $T$ in a non-trivial manner, in agreement with theoretical
expectations~\cite{bhattacharyya_05,stevenson_06,franz_07}, and that
this can in turn give rise to a non-monotonic $T-$dependence of the
dynamic length scale~\cite{kob_12}. This shows that it is insufficient
to characterize the CRRs just by means of a length scale. While direct
measurements of CRRs are still difficult~\cite{flenner_13}, the spatial
structure of the CRRs can also be probed indirectly.  Indeed, one can
study how the relaxation dynamics of the liquid depends on the size of
the system~\cite{berthier_12a} or, alternatively, how the dynamics is
influenced by the presence of a rigid wall~\cite{kob_12,hockey_14}.

In this latter type of study, the non-linear response of the
liquid to the external field of the rigid wall is used to probe
certain multi-point correlations. This is in fact just a special
case of a broader class of multi-point correlations functions that
can be measured by pinning a subset of particles of the liquid
at some instant of time and then measuring the evolution of the
remaining, i.e. unpinned, particles.  Recent investigations using
randomly pinned particles have indeed revealed static and dynamic
correlations whose associated length scales grow appreciably with
decreasing temperature and hence give insight into the nature of the glass
transition~\cite{kim_03,bouchaud_04,biroli_08,cammarota_12,cammarota_13,karmakar_13,hockey_12,kob_13}.
However, at present it is not really understood how the presence
of such pinned particles affects the relaxation dynamics in
a quantitative manner and to what extent this influence can be
captured by theoretical approaches. Since certain theories of the
glass-transition, such as the ``random first order transition''(RFOT)
theory~\cite{kirkpatrick_89,biroli_bouchaud_assessment,lubchenko_07,kirkpatrick_14},
make an intimate connection between the growing {\it dynamic} length
scales and an underlying {\it static} length scale, it is important
to obtain an accurate understanding of this dynamics so that it can be
compared with the static order.

In order to advance on this topic we present in the following extensive
simulation results on how the relaxation dynamics of a glass-forming
liquid is affected by the presence of pinned particles and how
thermodynamic theories can rationalize these findings. Such a study allows us to gain insight into the nature of the CRRs in the bulk around and below the dynamic crossover---a temperature regime seldom explored in computer simulations.

The system we consider is a 50:50 binary mixture of harmonic
spheres~\cite{ohern_02} of diameter ratio 1.4 at constant density
$\rho=0.675$ (more details are given in the SI). This system has been
shown to be a good glass-former, i.e.~it does not show any sign of
crystallization at the temperatures we consider here. To give the
relevant temperature scales of this model we recall that its onset
temperature is around $T_{\rm on} \approx 12$~\cite{flenner_jcp13} and
its mode-coupling theory (MCT) temperature $T_{\rm MCT} \approx 5.2$~\cite{kob_12}.
All numbers are expressed in appropriate reduced units (see SI). The
number of particles we consider is 20000 for studying the relaxation
dynamics and 1000 for the calculation of the variance $\chi_4(t)$
of the overlap correlation function. The pinning of the particles has
been done as in Ref.~\cite{kob_13}, i.e. the arrangement of the pinned
particles is uniform with a well characterized distance between them.
More details on this and the simulations can be found in the SI.

\begin{figure}[t]
\includegraphics[width=85mm,clip]{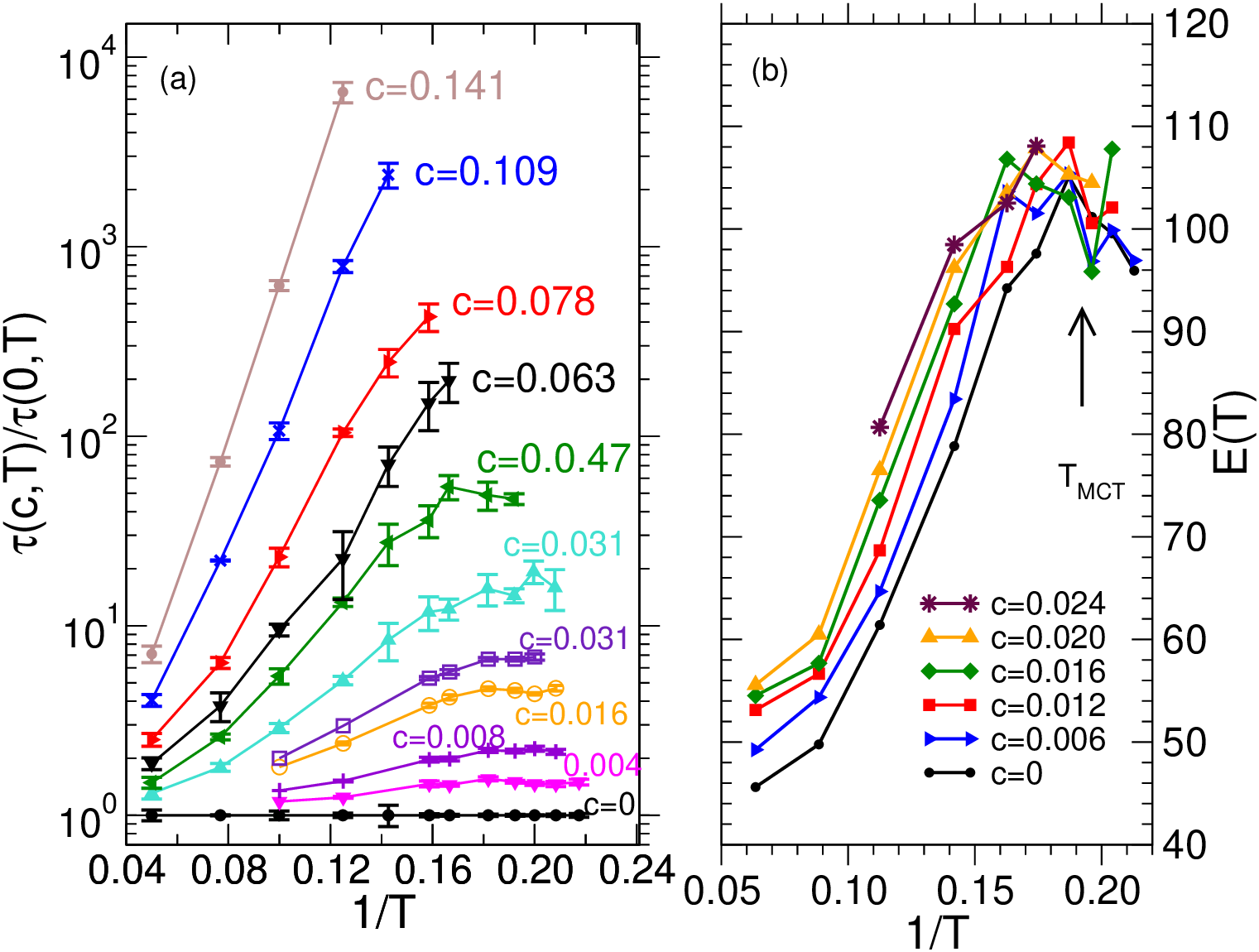}
\caption{a) Arrhenius plot of the $\alpha-$relaxation time $\tau(c,T)$,
normalized by the relaxation time of the bulk, for different
concentrations of pinned particles. The data for $c \leq  0.031$ is for
$N=20000$ particles the other for $N=1024$. b) $T-$dependence of the
activation energy $E(T)$ for different values of $c$ for $N=20000$.}
\label{fig1_tau_vs_T}
\end{figure}

To characterize the relaxation dynamics we have calculated
the self intermediate scattering function $F_s(q,t)$ using as
wave-vector $q$ the position of the first peak in the static
structure factor, i.e. $q=5.52$. As documented well in the
literature~\cite{kim_03,scheidler_04,kim_10,berthier_12}, the relaxation
dynamics slows down quickly if the concentration $c$ of pinned particles
increases (Fig. 1 in SI). In Fig.~\ref{fig1_tau_vs_T}a we show an
Arrhenius plot of the relaxation time $\tau(c,T)$, normalized by the bulk
value $\tau(0,T)$, for different values of $c$. Here we have defined
the relaxation time by the condition that $F_s(q,\tau)=e^{-1}$. Also
we mention that in the following we will always consider the larger
particles but we note that the relaxation dynamics of the small particles
is qualitatively very similar.

From the figure we recognize that, for a fixed value of $c$, at high and
intermediate temperatures the {\it normalized} relaxation time increases
with decreasing $T$ and then becomes basically flat, i.e. $\tau(c,T)$
tracks $\tau(0,T)$. This can also be clearly seen from the $T-$dependence
of the activation barrier, $E(T) = \frac{d\log(\tau)}{d(1/T)}$,
which at low $T$ becomes essentially constant and independent of $c$
(Fig.~\ref{fig1_tau_vs_T}b).  This change in the relaxation dynamic
strongly resembles the dynamic crossover observed experimentally in
several supercooled liquids~\cite{stickel_96,casalini_04}.  For small
$c$ this crossover occurs at around $T=5.5$, i.e. slightly above the
value of $T_{\rm MCT}$ of the bulk and as $c$ is further increased,
the crossover slightly shifts to higher temperatures.  These results
indicate that the dynamic response of the liquid to random pinning changes
qualitatively around the ($c-$dependent) dynamic crossover: The dynamics at
high and intermediate temperatures is increasingly affected by the pinned
particles as $T$ is decreased, whereas the one at low $T$ is only slowed
down (with respect to the bulk) by a constant factor. We emphasize
that the $T-c-$range we are probing here is far away from the
Kauzmann-line $T_K(c)$ investigated in Refs.~\cite{cammarota_12,kob_13}
and at which $\tau(c,T)/\tau(0,T)$ can be expected to diverge.

Such a change of the dynamics can be rationalized within the framework
of RFOT~\cite{bhattacharyya_05,stevenson_06}: We assume that the CRRs
are composed of a compact core, whose size grows with decreasing $T$,
surrounded by a ``halo'' consisting of string-like excitations connected
to this core. The size of these excitations grows if $T$ is lowered
towards $T_{\rm MCT}$ but once $T$ is well below $T_{\rm MCT}$, these
strings are no longer relevant for the relaxation of the system and
hence the CRRs consists only of the central core. Thus the increase in
$\tau(c,T)/\tau(0,T)$ we observe at intermediate $T$ can be explained
by the fact that the average length of the string-like excitations
is reduced because of the presence of the pinned particles and as a
consequence the dynamics slows down faster than $\tau(0,T)$. However,
once $T$ is below $T_{\rm MCT}$ the effective size of the CRRs shrinks
 and hence $\tau(c,T)$ tracks $\tau(0,T)$, i.e. $\tau(c,T)/\tau(0,T)$
becomes a constant. This interpretation of the data is also compatible
with the results of Refs.~\cite{kob_12,hockey_14} for which it has
been found that the dynamic length scale shows around $T_{\rm MCT}$
a local maximum/saturation.

\begin{figure} 
\includegraphics[width=85mm,clip]{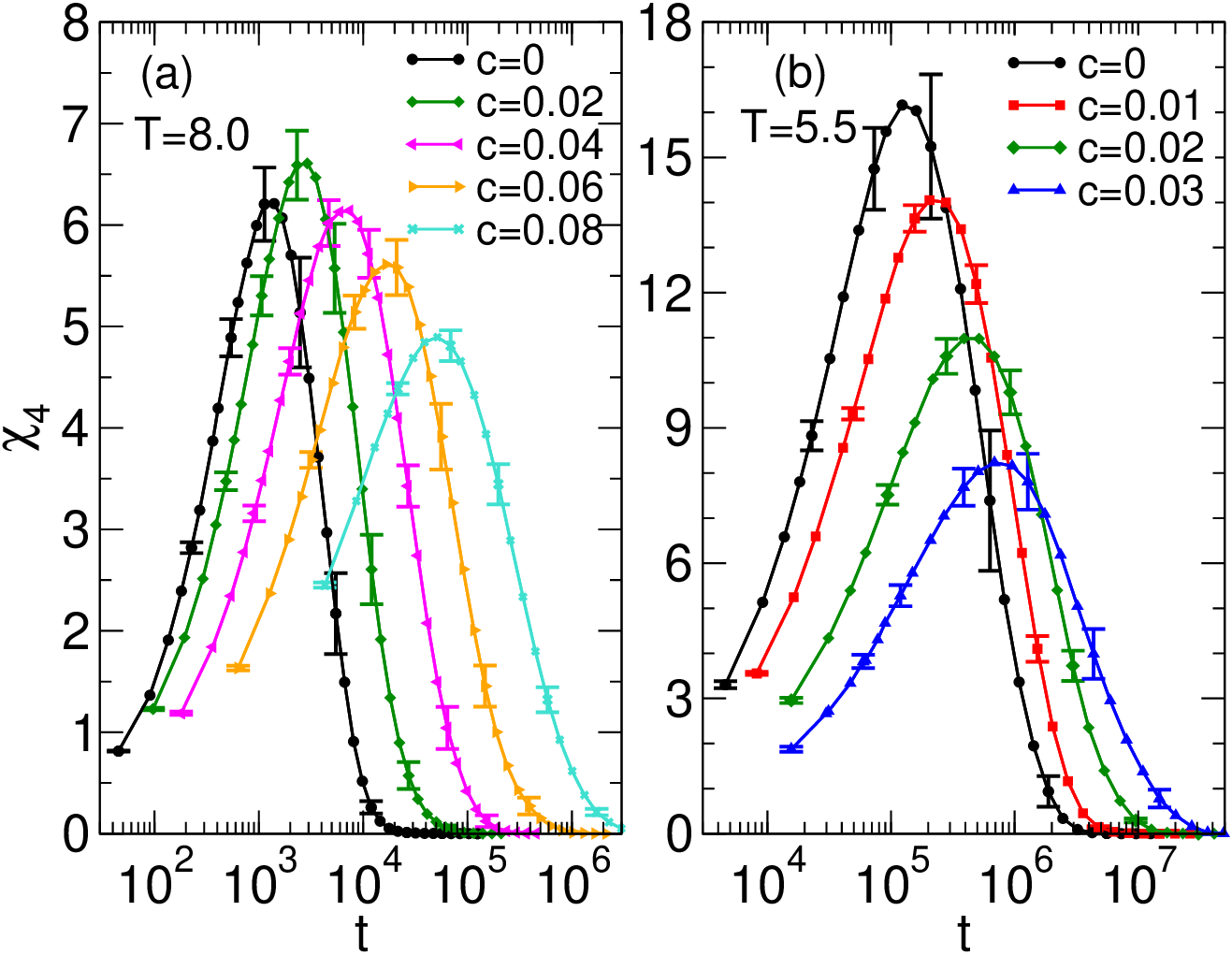}
\caption{Time dependence of $\chi_4$ for different values of $c$ and
$T$. a) $T=8.0$, b) $T=5.5$. Note the different scales of the ordinate in 
the two panels.}
\label{fig2_chi4}
\end{figure}

To shed more light on the nature of the relaxation dynamics we have
characterized its degree of cooperativity. For this we use the dynamic
susceptibility $\chi_4(t)$, i.e.~the variance of a time correlation
function that characterizes the relaxation dynamics~\cite{berthier_07}. In
practice, we have used the self-overlap $\phi_s(t)$ but it can be expected
that the self intermediate scattering function gives qualitatively the
same results~\cite{flenner_jcp13}. Details are given in the SI.

In Fig.~\ref{fig2_chi4} we show the $t-$dependence of $\chi_4$
for different values of $c$ and $T$. As usual, this quantity
shows a maximum the height of which is related to the degree of
cooperativity of the relaxation process~\cite{berthier_07}. At
intermediate temperatures, see Fig.~\ref{fig2_chi4}a, the height
of the peak depends only weakly on $c$ and its amplitude remains
rather small. This shows that at this $T$ the cooperativity is not
very pronounced and only marginally affected by pinning, in agreement
with the results shown in Fig.~\ref{fig1_tau_vs_T}. At temperatures
close to $T_\textrm{MCT}$, Fig.~\ref{fig2_chi4}b, the $c-$dependence
of $\chi_4$ is much more pronounced and we see that the cooperativity
{\it decreases} with increasing $c$. This result is reasonable since,
as discussed above, the pinned particles reduce the effective size
of the CRRs~\cite{footnote1}. If we denote by $\xi(c,T)$ the extent
of the CRRs for pinning concentration $c$ and temperature $T$, we can
expect that $\xi(c,T)$ is bounded from above by $c^{-1/3}$, i.e.~by the
typical distance of the pinned particles. If we assume that the CRRs are
compact, one expects that $\chi_4^\star$ should scale as $\xi^3 \propto
c^{-1}$~\cite{footnote2} {Note that this estimate holds also if the shape
of the CRRs is not spherical.}. Our data, shown in Fig.~3 of the SI,
shows that $\chi_4^\star$ does indeed decrease with growing $c$ and that
there is an upper bound to its value. However, this upper bound is not as
expected described by a $c^{-1}-$dependence but rather by a $c^{-2/3}$
law. This suggests that the CRRs have a fractal, rather than compact,
structure.  Finally we mention that the reduced cooperativity as $c$ is
increased explains why the $T-$dependence of the relaxation times becomes
Arrhenius-like at high $c$, see Fig.~\ref{fig1_tau_vs_T}, a result that is
also in qualitative agreement with the ones from Ref.~\cite{cavagna_12}.

\begin{figure}
\includegraphics[width=85mm,clip]{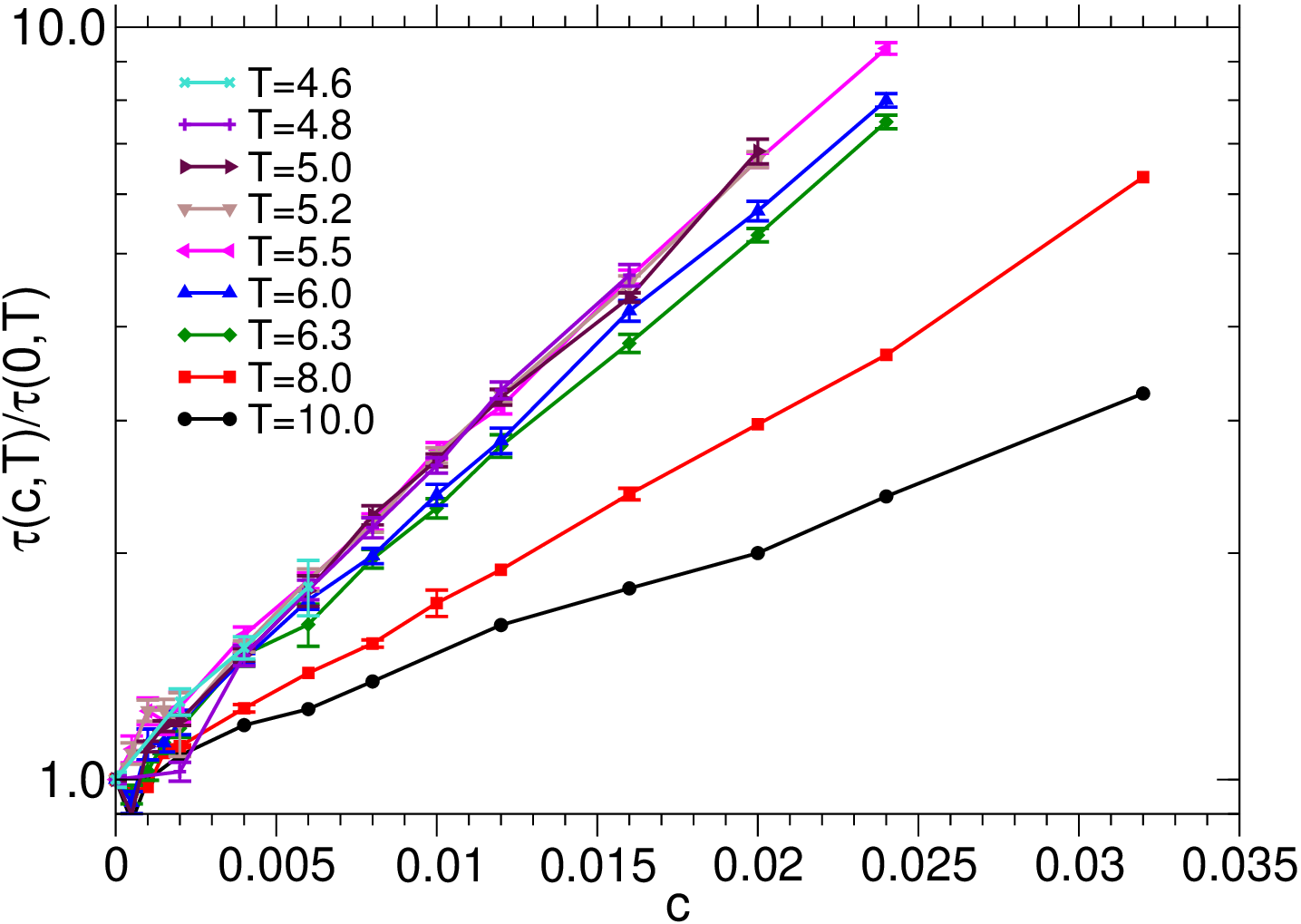}
\caption{$\alpha-$relaxation time $\tau(c,T)$, normalized by the
relaxation time of the bulk, as a function of concentration of pinned
particles.}
\label{fig3_tau_vs_c}
\end{figure}

\begin{figure}
\includegraphics[width=85mm,clip]{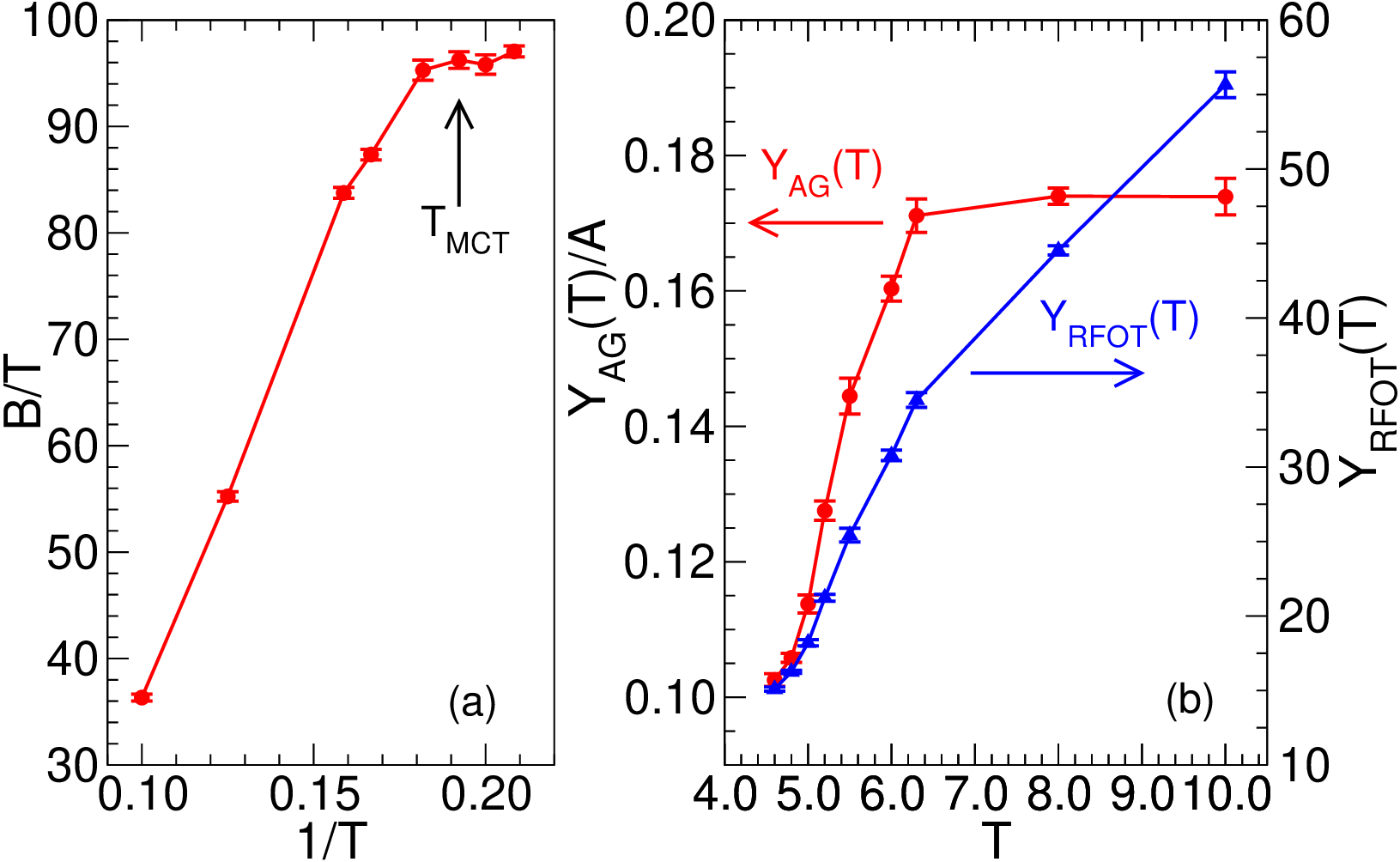}
\caption{
a) $T-$dependence of the slope of the curves in Fig.~\ref{fig3_tau_vs_c}.
The $T-$dependence of this slope changes significantly at a temperature
that is close to the mode-coupling temperature of the system.  b)
$T-$dependence of the ratios $Y^{\rm AG}(T)/A$ (circles) and $Y^{\rm
RFOT}(T)$ (trangles) from the Adam-Gibbs theory and the RFOT theory,
respectively.}
\label{fig4_Y_T_AG_RFOT}
\end{figure}

We now turn our attention to the $c-$dependence of the relaxation time. In
Fig.~\ref{fig3_tau_vs_c} we show the ratio $\tau(c,T)/\tau(0,T)$ as a
function of $c$ for different temperatures. Within the accuracy of the
data, $\tau(c,T)$ has an exponential dependence on $c$~\cite{footnote3}
\cite{footnote4}. We note that the prefactor in the exponent,
i.e. the slope of the curves, shows a significant $T-$dependence at
high and intermediate temperatures, but depends only weakly on $T$
at low temperatures (see Fig.~\ref{fig4_Y_T_AG_RFOT}a).  Below we will
discuss this $T-$dependence in the context of thermodynamic theories
for the glass-transition. This pronounced change in the $T-$dependence
occurs at a temperature that is close to $T_{\rm MCT}$, thus giving
further indication that at this temperature the nature of the relaxation
dynamics changes significantly. Furthermore we mention that we have also
determined the $c-$dependence of the infinite-time overlap (see Fig.~SI4
in SI for details) and found that this static quantity shows a linear
dependence on $c$ with a slope that is basically independent of $T$. This
shows that non-linearities in the dynamics are found in a concentration
regime over which the static response of the liquid is linear.

The exponential $c-$dependence of $\tau(c,T)$ indicates that our
simulation probe a concentration regime where pinning linearly affects
the activation energy, rather than the bare relaxation times.
This ``non-local'' effect can be easily rationalized within either the
Adam-Gibbs or the RFOT theory. To see this we set the Boltzmann
constant $k_B=1$ and we write the relaxation times as\\[-8mm]

\begin{equation}
\tau(c,T)= \tau_0(c) \exp\left[ E(c,T)/T \right] \quad ,
\label{eq1}
\end{equation}

\noindent
where $\tau_0(c)$ is a prefactor that we will assume to be independent
of $c$ and $T$, and $E(c,T)$ can be interpreted as an effective $c-$and
$T-$dependent activation energy. Expanding $E(c,T)$ around $c=0$, we can
write $E(c,T) \approx E(0,T)+ B(T)c$ from which one obtains immediately
an exponential $c$-dependence\\[-8mm]

\begin{equation}
\frac{\tau(c,T)}{\tau(0,T)} = \exp[cB(T)/T] \quad .
\label{eq2}
\end{equation}

Within the Adam-Gibbs and RFOT frameworks the bulk relaxation time,
$\tau(0,T)=\tau_0 \exp[E(0,T)/T]$, can be connected to the
thermodynamic properties of the system. In the following we will
generalize this connection to the pinned particles case. For this we
assume that for small values of $c$ the pinning leads to a reduction
of the configurational entropy $s_c(c,T)$ via~\cite{cammarota_12}\\[-8mm]

\begin{equation}
s_c(c,T)=s_c(T)-Y(T)c \quad .
\label{eq0}
\end{equation}

Within the Adam-Gibbs theory the activation barrier can thus be written as\\[-8mm]

\begin{equation}
E_{\rm AG}(c,T)= A/s_c(c,T) \quad ,
\label{eq3}
\end{equation}

\noindent
where $A$ is a constant, from which one immediately finds\\[-8mm]

\begin{equation}
B_{\rm AG}(T)=Y_{\rm AG}T^2A^{-1}\ln^2(\tau(0,T)/\tau_0) \quad .
\label{eq4}
\end{equation}

Generalizing the expression proposed by the RFOT theory for
$E(0,T)$~\cite{kirkpatrick_89,rabochiy_13} to $c>0$ one can expect that\\[-8mm]

\begin{equation}
A_{\rm RFOT}(c,T)= \frac{3\pi (1.85)^2T}{s_c(c,T)} \approx \frac{32T}{s_c(c,T)}\quad .
\label{eq5}
\end{equation}

\noindent
Although in the framework of RFOT other expressions for $E(0,T)$ have
been discussed in the literature~\cite{rabochiy_13}, Eq.~(\ref{eq5})
is a reasonable approximation. Together with Eq.~(\ref{eq0}) this leads 
to\\[-8mm]

\begin{equation}
B_{\rm RFOT}(T)= \frac{T Y_{\rm RFOT}(T)}{32} \ln^2(\tau(0,T)/\tau_0) \quad .
\label{eq6}
\end{equation}

\noindent
Note that in Eqs.~(\ref{eq4}) and (\ref{eq6}) the quantity $Y(T)$
has a label that refers to the theory considered. 
In fact, the actual definition of the configurational entropy (and hence the meaning of
$Y(T)$ as defined in Eq.~(\ref{eq0})) depends on the theory. These two
expressions, together with Eq.~(\ref{eq2}), are qualitatively consistent
with the observed $c-$dependence of the relaxation times.

Since the Adam-Gibbs theory is purely phenomenological, it is not
possible to predict {\it a priori} the $T-$dependence of $Y_{\rm AG}$.
In contrast, it should in principle be possible to determine $Y_{\rm
RFOT}$ with RFOT, but so far no explicit expression is known for
this quantity. On the other hand, we can use our simulation data
for $\tau(c,T)/\tau(0,T)$ and Eqs.~(\ref{eq4}) and (\ref{eq6})
to extract the $T-$dependence of the quantities $Y_{\rm AG}(T)/A$
and $Y_{\rm RFOT}(T)$, respectively and the results are shown in
Fig.~\ref{fig4_Y_T_AG_RFOT}b. From this figure we see that the ratio
$Y_{\rm AG}(T)/A$ is a constant at intermediate and high temperatures,
decreases rapidly on approaching $T_{\rm MCT}\approx 5.2$ before it
starts to level off at even lower temperatures. Qualitatively the
same $T-$dependence is found for $Y_{\rm RFOT}$, since basically this
quantity differs from $Y_{\rm AG}(T)$ just by a factor of $T$. Thus,
in this latter case, a linear $T$-dependence is observed at intermediate
and high temperatures.

Thus, within the framework of AG or RFOT, the physical origin of the
decrease of $Y(T)$ with decreasing $T$ is that the size of the CRRs
is increasing: At high $T$, pinning one more particle is more likely
to block the dynamics of an additional CRR, since there are many of
them and they are not extended. Hence, increasing $c$ changes the
configurational entropy significantly, i.e.~$Y(T)$ is large.  However,
at low $T$ additional pinned particles are likely to be found in regions
that are already very slow (since there are only few CRRs and they are
extended) and hence the impact of pinning on $s_c$ is rather weak,
i.e.~$Y(T)$ is small.  The fact that the slope $B/T$ shows around
$T_{\rm MCT}$ a marked change in its $T-$dependence is thus directly
related to a change in the $T-$dependence of the dynamic length scales,
a change that can be rationalized, at least qualitatively, within
RFOT~\cite{kirkpatrick_89,lubchenko_07}.

In summary, the present simulation results show that studying the
dynamics of glass-forming liquids with pinned particles allows to gain
insight into the nature of the relaxation process {\it in the bulk},
such as the $T-$dependence of the structure of the CRRs. Although
this approach remains indirect, it is in principle realizable in
experiments using optical tweezers on colloidal suspensions or in
granular materials. Therefore it will allow to probe novel details about
the dynamics in glass-forming liquids and thus pose a new challenge to
microscopic theories like RFOT to rationalize the $T-$dependence of $Y(T)$
on a quantitative level. All this will therefore help in the quest to
find the correct theory for the glass-transition.

Acknowledgments: We thank G. Biroli for useful discussions.  This work
was realized with the support of HPC@LR  Center of Competence in
High Performance Computing of Languedoc-Roussillon (France).  W. Kob
acknowledges support from the Institut Universitaire de France.

% @Keiji Watanabe and Hajime Tanaka prl 2008 discuss a similar connection
% betwen size of cristalline regions and relaxation times (but no pinning)

% @see also tanaka et al NatMat 2010

%\appendix

\newpage

\section{Supplementary Information}

\renewcommand{\figurename}{FIG. SI}
\setcounter{figure}{0}

{\bf Model and details of the simulation}

The system we study is a 50:50 mixture of elastic spheres~[1]. Both
type of particles have the same mass $m$ and the interaction between a
particle of type $i$ and $j$ is given by\\[-8mm]

\begin{equation}
V(r_{ij}) = \frac{\epsilon}{2} (1-r_{ij}/\sigma_{ij})^2 \quad 
\label{eq7}
\end{equation}

\noindent
if $r_{ij} < \sigma_{ij}$ and zero otherwise. Here $\sigma_{11}=1.0$,
$\sigma_{12}=1.2$ and $\sigma_{22}=1.4$. In the following we will use
$\sigma_{11}$ as the unit of distance, $\sqrt{m\sigma_{11}/\epsilon}$
as the unit of time and $10^{-4}\epsilon$ as unit of energy, setting
the Boltzmann constant $k_B=1.0$.  The static and dynamic properties of
the system have been obtained via molecular dynamics using the Verlet
algorithm with a time step of 0.01. The simulations for the system with
20000 particles have been carried out using the LAMMPS package~[2]. In
order to improve the statics we have typically simulated 8 independent
samples. The longest simulations were $3\cdot 10^9$ time steps. The
simulations for intermediate and high values of $c$ have been done with
1024 particles using typically 4 samples. The data for the four-point
susceptibiliity $\chi_4(t)$ has been obtained for systems of $N=1000$
particles using typically from 6 to 20 samples depending on temperature and pinned particles concentration.\\[5mm]

{\bf Relaxation times}

\begin{figure}[tbh]
\includegraphics[width=85mm,clip]{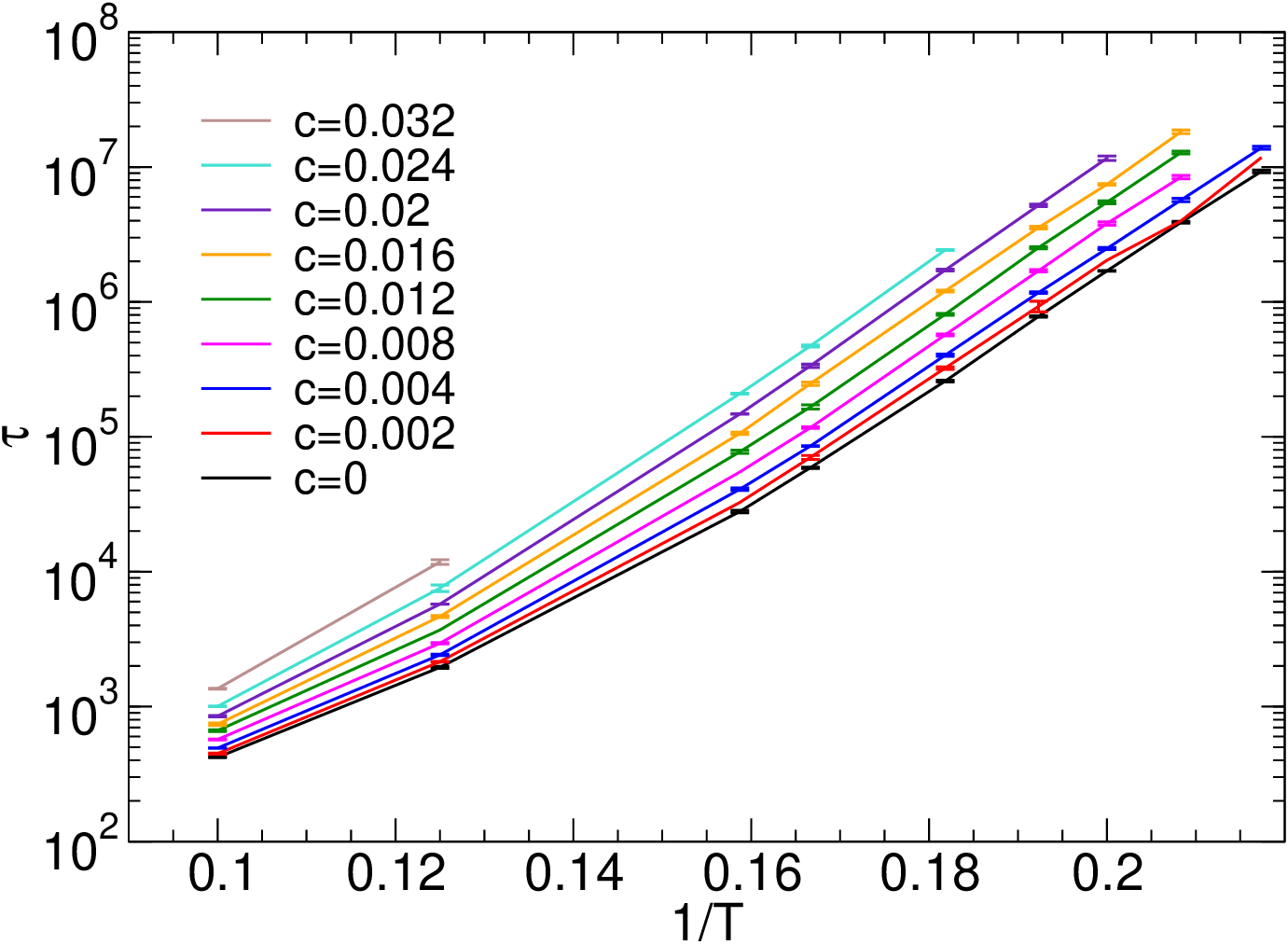}
\caption{Arrhenius plot of the relaxation time for different values of $c$.}
\label{SI_fig1}
\end{figure}

We have determined the relaxation time $\tau(c,T)$ from the self
intermediate scattering function $F_s(q,t)$ by requiring that at $\tau$
this correlator has decayed to $e^{-1}$. The wave-vector used was
$q=5.52$, i.e.~close to the location of the maximum in the static structure
factor. The temperature dependence of $\tau$ is shown in Fig.~SI\ref{SI_fig1}
for different values of $c$ and $N=20000$. As can be seen immediately,
an increase of $c$ leads to a significant increase of $\tau$.

\begin{figure}[b]
\includegraphics[width=75mm,clip]{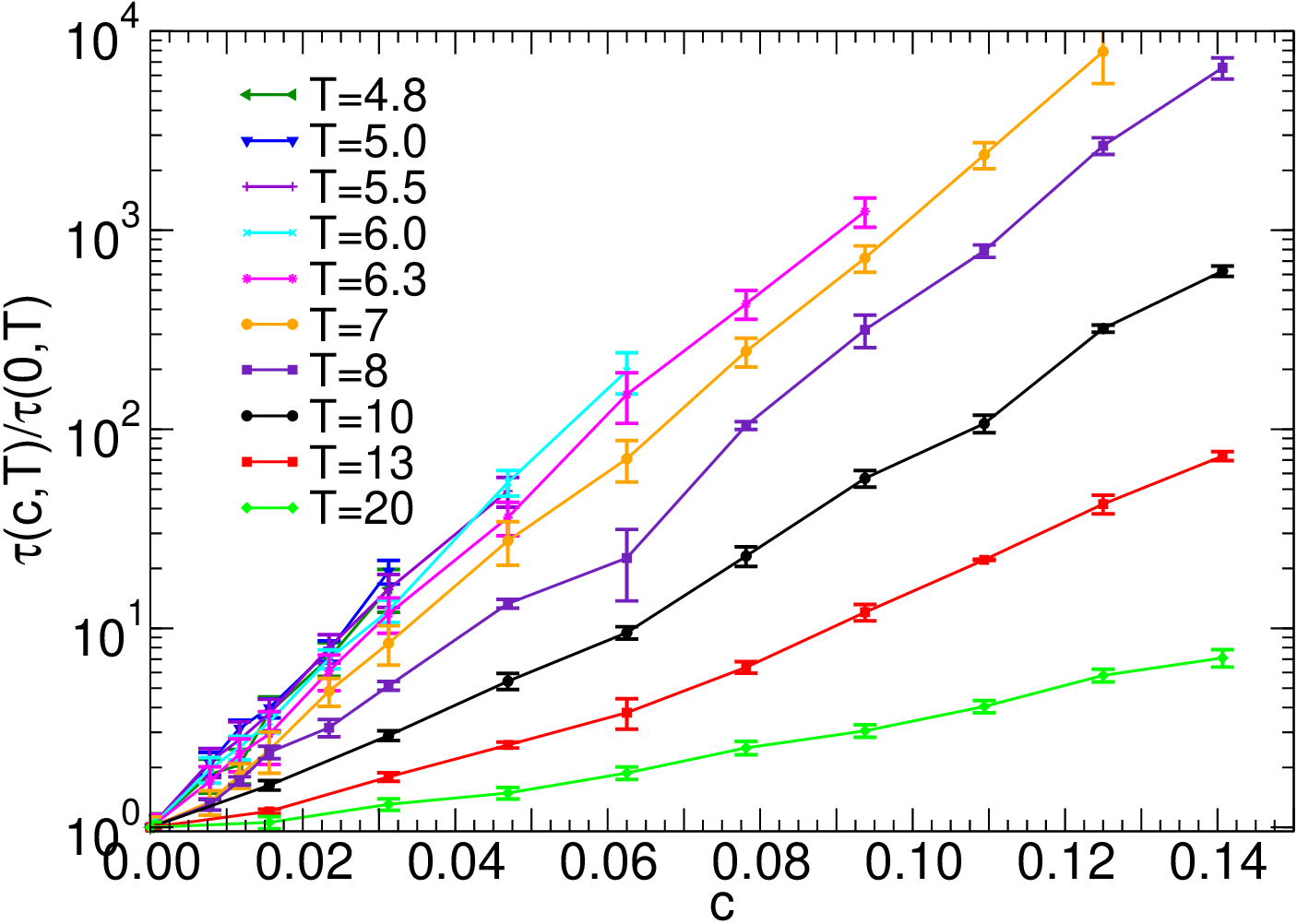}
\caption{Normalized relaxation time as a function of $c$ for $N=1024$.}
\label{SI_fig3}
\end{figure}

In the main text we have shown, see Fig.~\ref{fig3_tau_vs_c}, that the
relaxation times show an exponential dependence on $T$ if $c$ is small. In
Fig.~SI\ref{SI_fig3} we show the analogous plot for a system of $N=1024$
particles for which we have been able to follow the relaxation dynamics to
higher concentrations. From this graph we recognize that the exponential
dependence hold also for  values that are significantly higher than the
ones shown in Fig.~\ref{fig3_tau_vs_c}. We emphasize, however, that the
maximum shown value of the concentration, $c=0.14$ is still significantly
smaller than the critical value at which one expects the ideal glass
transition to occur, which, for $T=6.3$ is around 0.19~[3]
and where a super-exponential $c-$dependence can be expected.\\[0mm]

{\bf Overlap and $\chi_4(t)$}

To characterize the cooperativity of the dynamics and in order to make
contact with previous work, we have studied the fluctuations of the
time-dependent overlap function $Q_s(t)$ which is defined as

\begin{equation}
Q_s(t) = N^{-1} \sum_i \Theta(|\vec{r}_i(t)-\vec{r}_i(0)| - a) \quad .
\end{equation}

\noindent
Here $\Theta$ is the Heavyside function and $a$ is a constant which is
typically a fraction of the interparticle diamater (in our case $a=0.3$).
The four-point susceptibility $\chi_4(t)$ was then evaluated as the
variance of the overlap function

\begin{equation}
\chi_4(t) = [\langle Q_s(t)^2 \rangle - \langle Q_s(t) \rangle^2] \quad ,
\end{equation}

\noindent
where $\langle \dots \rangle$ denotes a thermal average for a fixed
realization of pinned particles and $[ \dots ]$ is the average over
the disorder.  Note that our definition of $\chi_4$ does not account
for sample-to-sample fluctuations, which would be included in the quantity

\begin{equation}
\chi_4^\textrm{full}(t) = [\langle Q_s(t)^2 \rangle] - [\langle Q_s(t) \rangle]^2 
\quad .
\end{equation}

\begin{figure}[tb] 
\includegraphics[width=85mm,clip]{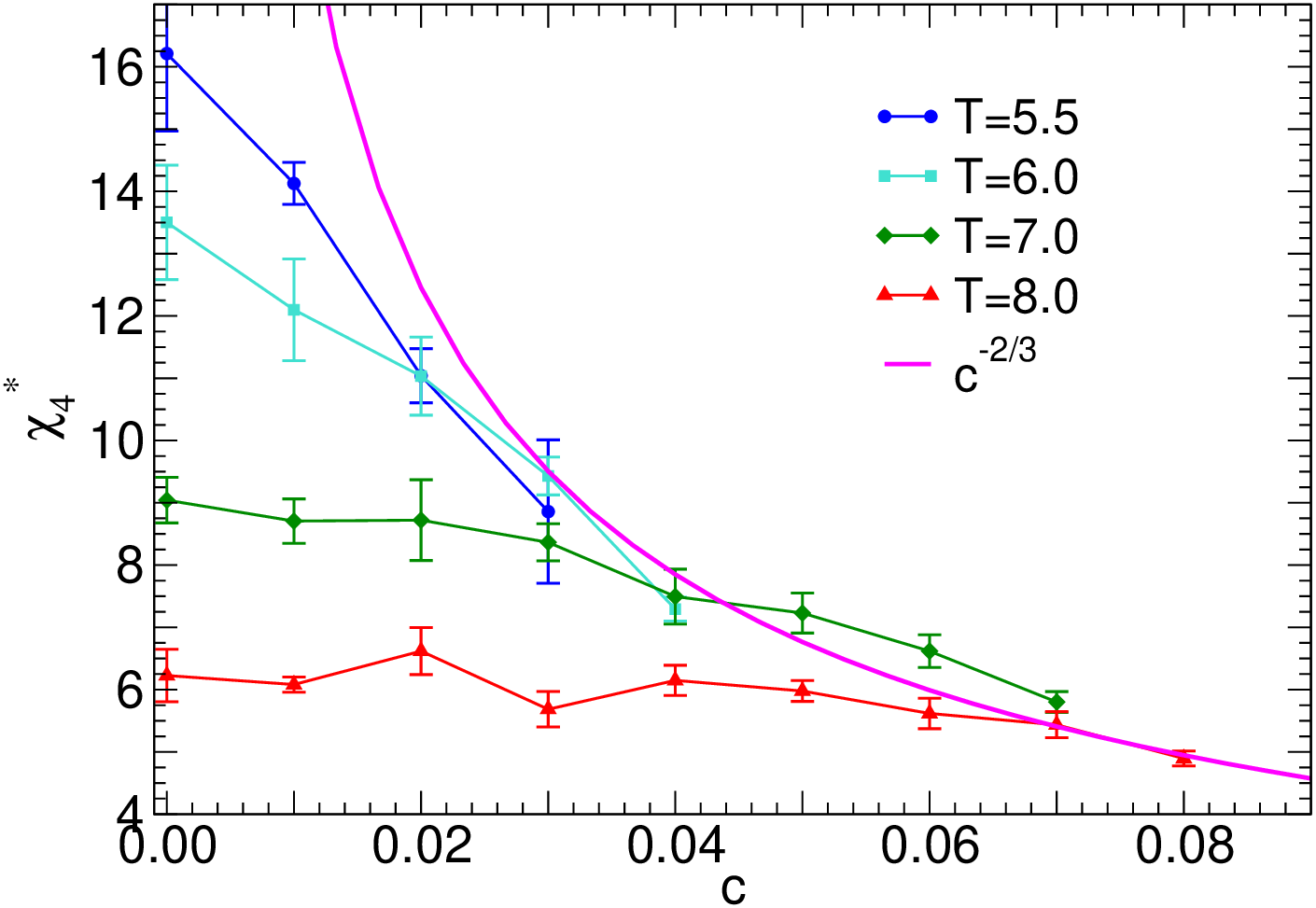}
\caption{Height of peak in $\chi_4(t,T)$ as a function of $c$ for different 
temperatures. Also show is the function $c^{-2/3}$ which makes an envelope to the data.}
\label{SI_fig2}
\end{figure}

We have found that the contribution of sample-to-sample fluctuations
are small compared to the thermal one, at least in the concentration
regime we have explored in this work, and for system sizes of $N=1000$
particles. The behavior might be different, however, at larger $c$
and on approaching the putative ideal glass transition.  We remark
that $\chi_4^\textrm{full}(t)$ was used in Ref.~[4]. The maximum value
of $\chi_4(t)$, $\chi_4^\star$, is shown in Fig.~SI\ref{SI_fig2} as a
function of $c$ for different $T$'s. From the figure one recognizes that
the data is bounded by the function $c^{-\alpha}$ with $\alpha \approx
2/3$, giving evidence for the fractal nature of the CRRs.

\begin{figure}[tb]
\includegraphics[width=75mm,clip]{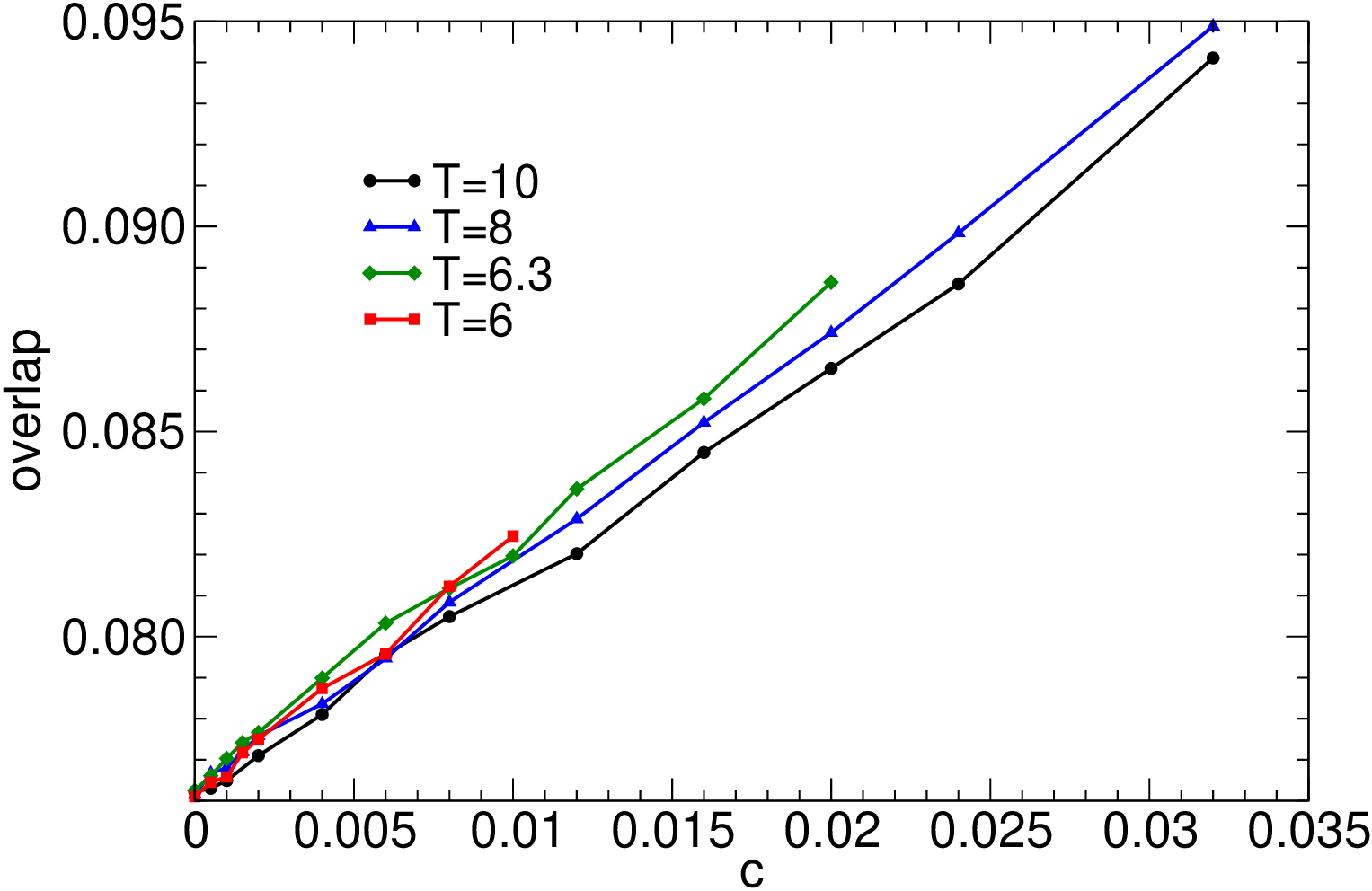}
\caption{Overlap as a function of $c$ for different temperatures.}
\label{SI_fig4}
\end{figure}

We have also considered the collective overlap function $Q_c(t)$, defined as

\begin{equation}
Q_c(t) = N^{-1} \sum_i \sum_j \Theta(|\vec{r}_i(t)-\vec{r}_j(0)| - a) \quad .
\end{equation}

\noindent
This quantity shows a time dependence that is qualitatively similar to
the one of the intermediate scattering function $F_s(q,t)$ and $Q_s(t)$,
but at long times it decays on a plateau with a non-zero height. This
height is the long time overlap and it characterizes how similar two
completely independent configuations are. The $c-$dependenc of this
overlap is shown in Fig.~SI\ref{SI_fig4} and one see that it is a linear
function in $c$, as expected at low $c$ where the zone of influence of
a given pinning particle is independent of the other frozen particles.\\[8mm]

%\begin{thebibliography}{2}

{\bf References}

%\bibitem{ohern_02_app}
[1] C. S. O'Hern, S. A. Langer, A. J. Liu, and S. R. Nagel,
Phys. Rev. Lett. {\bf 88}, 075507 (2002).

%\bibitem{lammps_app} 
[2] S. Plimpton,
J. Comp. Phys. {\bf 117}, 1 (1995).
%http://lammps.sandia.gov

%\bibitem{kob_13_app}
[3] W. Kob and L. Berthier,
Phys. Rev. Lett. {\bf 110}, 245702 (2013).

%\bibitem{fullerton_13_app}
[4] R. L. Jack and C. J. Fullerton 
Phys. Rev. E {\bf 88}, 042304 (2013).

%\end{thebibliography}


\begin{thebibliography}{99}
\bibitem{footnote2}
Note that this estimate holds also if the shape
of the CRRs is not spherical.
\bibitem{footnote3}
The data shown is for $N=20000$ particles and, due to the needed
computational effort, we have only data for up to $c=0.03$. However, we
have also done simulations for $N=1024$ particles and for this smaller
system size we have been able to extent the $c-$range up to $c=0.14$. The
same exponential $c-$dependence has been found, thus showing that it
holds up to relatively high concentrations. See Fig.~SI2 in the SI for
more details.
\bibitem{footnote4}
We mention that we have also carried out simulations of the same system
using a Langevin dynamics. Although the absolute time scales for the
relaxation are different from the one of the Newtonian dynamics, the
{\it normalized} times are identical, thus showing that the exponential
$c-$dependence is independent of the microscopic dynamics.
\end{thebibliography}
\end{document}